\documentclass[journal=jacsat,manuscript=article]{achemso}

\ifnum \value{page}=1
\usepackage{chemformula} 
\usepackage[T1]{fontenc} 

\author{Asl{\i} \"Ozt\"urk Kiraz}
\affiliation{Computational Physics Laboratory, Department of Physics, Pamukkale University, Denizli, Turkey.}
\author{Ali Ba{\u g}c{\i}}
\affiliation{Computational Physics Laboratory, Department of Physics, Pamukkale University, Denizli, Turkey.}
\author{Philip E. Hoggan}
\affiliation{Institut Pascal, UMR 6602 CNRS, BP 80026, 63178 Aubiere Cedex, France.}
\email{philip.hoggan@uca.fr}

\title[An \textsf{achemso} demo]{Quantum Monte Carlo method for metal-film catalysis: water addition to carbon monoxide adsorbed on Pt/Al(111), a route to hydrogen.}


\begin{document}


\begin{abstract}

Hydrogen production as a clean, sustainable replacement for fossil fuels is gathering pace. Doubling the capacity of Paris-CDG airport has been halted, even with the upcoming Olympic Games, until hydrogen-powered planes can be used.
It is thus timely to work on catalytic selective hydrogen production and optimise catalyst structure. Over 90 \% of all chemical manufacture uses a solid catalyst.
This work describes the adsorption of carbon-monoxide (CO) on platinum thin films, supported by cheap Al(111). CO reacts with water to produce hydrogen (water-gas shift).
Quantum Monte Carlo methods are the only ones accurate enough to investigate the early steps of this catalysed reaction at close-packed Pt/Al(111).

Many chemical reactions involve bond-dissociation. This process is often the key to rate-limiting reaction steps at solid surfaces.
Since bond-breaking is poorly described by Hartree-Fock and DFT methods, our embedded active site approach is used. This work demonstrates a novel Quantum Monte Carlo (QMC) methodology.

The water-gas shift reaction step studied is water addition to CO pre-adsorbed on a Pt-monolayer supported by Al(111). The water molecule is only partially dissociated.  Its oxygen atom binds to CO giving adsorbed COOH and Pt-H. This concerted addition is rate-limiting. In subsequent steps, the adsorbed formate species (with acidic hydrogen) decomposes to carbon dioxide and, after proton migration to Pt-H, molecular hydrogen is obtained as the the clean product.



 The QMC activation barrier found is 64.8 $\pm$ 1.5 kJ/mol \cite{shar}. Thus, QMC is shown to be encouraging for investigating similar catalytic systems.

\end{abstract}

\section{Introduction}

\hskip5mm The author and his collaborators have consistently used Quantum Monte Carlo methods at chemical accuracy (1 kcal/mol=4.2 kJ/mol) for activation barriers of reactions at metal catalyst surfaces, since publishing test benchmarks on hydrogen dissociation on Cu(111) in 2015. Progress has been made in controlling error sources with the Casino QMC code for periodic systems by K. Doblhoff-Dier and P. E. Hoggan \cite{kdd1,kdd2}. Our early work on adsorbed CO is in \cite{hog1}.

The latter has worked on selective hydrogen production at the Pt(111) surface using water-gas shift. A recent publication gives the rate-limiting barrier for water addition to Pt(111), in this context to within 1kJ/mol \cite{hjcp,shar}. This benchmark uses a Multi-reference (MRCI) trial wave-function, suited to bond breaking and formation and an embedded active site approach to the catalyst, allowing high-level input to be used for a small molecular active site, before the periodic QMC. These QMC calculations took 50Mh (PRACE {\it vide infra.}).

Other applications to catalysis and weak interactions of accuracy close to 1 kJ/mol are rare and often used to calibrate DFT functionals for cheaper investigations \cite{kdd3, oakr}.

It clearly would be of interest to obtain valid information on the catalytic process with simpler wave-functions. We have since tested the use of single-determinant (ground state) wave-functions for the same system.

Here, we estimated the systematic error due to imperfect nodes supplied in these simpler trial wave-functions to 2.5 kJ/mol. In fact, the barrier is over-estimated since the transition state is strongly multi-reference \cite{hoggadv}. The order of magnitude of this error is already that obtained by comparing a single-determinant to MRCI input for the molecular part of the system \cite{molpro}.

In this new work, a short configuration state function (CSF) was used to study a different avenue towards selective hydrogen production on a metal catalyst.

This work continues previous studies that suggest a thin films of Pt on (cheap) Al(111) is effective catalysts for water attack on pre-adsorbed CO. Some indications of this method were already described in \cite{hoggadv}, showing the thin Pt-film is a much better catalyst than Pt-doped Al(111), as well as being more readily prepared, in practice.

Nowadays, stochastic methods are increasingly useful in
work on the challenges of real systems which require electron correlation accurately.
One such growth area for application is heterogeneous catalysis, in particular adsorbed reactions on metals.

This study focusses on the early stages of the water-gas shift reaction (wgs) which follows the overall equation:

CO + H$_2$O $\rightarrow $  CO$_2$ + H$_2$

This reaction was discovered in 1780 by Fontana. Here, the reaction is investigated on Pt(111). Much of the mechanistic work has postulated a redox mechanism, particularly for metals like copper which are known to be reducing agents \cite{mavr}. The alternative mechanism is associative, evidenced in this work. No change in the Pt-oxidation state is seen whilst an additive complex between CO and water is characterised. After several more elementary steps (not studied here), hydrogen (and CO$_2$) are produced from formate decomposition. A full study of this mechanism on Pt, including experiment, DFT and kinetics is proposed in the review \cite{mukin}.
This reference also collects previously published activation barriers for the CO oxidation involved.
These barriers cover a factor of two \cite{mukin}.
\newpage
Nevertheless, several new DFT functionals have emerged that improve performance compared to calorimetric measurements of adsorption energies and the approach continues to be very informative, including on nano-structured substrates and kinetics, when combined with kinetic Monte Carlo. See, for example \cite{Camp, Clay, Faj2020}.

In this work, we introduce the highly accurate Quantum Monte Carlo (QMC) approach for polyatomic active sites embedded in a periodic solid.


This theoretical study of a step in wgs, i.e. water addition to pre-adsorbed carbon monoxide in a hollow site in the Pt(111) plane of the catalyst develops novel QMC methodology. Our earlier benchmark \cite{hjcp} shows that single determinant work on the same system over-estimated the barrier by 2.5 kJ/mol, whereas that on the thin film of Pt(111) is estimated over 6 kJ/mol lower, because of a larger lattice parameter for Pt(111) stretched by the Al(111) substrate that gives the layer catalyst greater activity. This phenomenon is not only significant and opposed in sign to the systematic error related to imperfect wave-function nodes but is readily explained from the stretched Pt close-packed active catalyst layer.

\section{Theoretical Method:}

\hskip4mm The method of choice must scale well with system size, be able to efficiently use modern supercomputer facilities that are massively parallel and, above all, produce quantitative, accurate physical properties that are difficult to obtain otherwise. QMC benchmarks of activation barriers for reactions adsorbed on solid catalyst surfaces fit this description well.
Developing and applying QMC calculations on this scale require access to a supercomputer.

\subsection{The stages of our approach are as follows:}

1-Define the active site with 2 Pt-atoms, one receives H the other (with translation symmetry) describes the equilateral triangle motif in the (111) face and the hollow site where CO is adsorbed at the triangle's centre of mass: The atoms involved are Pt1, Pt2, CO and H$_2$O.

The TS input geometry, from QMC force constants is:(molecular active site, Bohr).

  Pt1,   0.0000     0.0000     0.0000 \hskip10mm

  Pt2,   2.6137     4.5271     0.0000

   C,   2.6137     3.0179     4.3000

   O,   2.6137     4.4344     6.5864

   O,   1.9000     0.0000     4.3000

   H,   0.0000     0.0000     3.0500

   H,   1.3155     0.0000     6.0126

2-Molpro uses this input to evaluate a Multi Reference Configuration Interaction (MRCI) wave-function, which is embedded in the Pt-lattice with a 2x2 atom face and 5 layers; ABABA. This defines a periodic slab. The 'active site' embedded in Pt-slab (without duplicating Pt-atoms) is expanded in plane waves, with periodic boundary conditions in k-space.


Embedding proceeds smoothly using a Green matrix addition of electronic potentials \cite{geo}.

3-The Slater determinants obtained from this wave-function are expressed in Slater-Jastrow form (as products with a generic Jastrow function).

4-This trial wave function with arbitrary parameters (polynomial coefficients of inter-particle distances in the Jastrow factor and Slater determinant weights) are optimised using Variational Monte Carlo (VMC).

The Quantum Monte Carlo approach uses statistical physics over a large population, comprising sets of instantaneous particle positions in co-ordinate space. They are often called 'walkers' (c.f. the one-dimensional random-walk. Random numbers actually serve to initialise the 'walkers' from the trial wave-function giving initial electron density/population).




5-This wave-function is then used to generate a population of 'walkers' (configurations) that are propagated in imaginary time during the second, diffusion step (DMC). DMC is carried out in the fixed-node approximation, which uses the nodes from the input trial wave-function. An updated overview of the VMC and DMC methods is given in \cite{cas2020}. These trial wave-functions can be optimised with a complex Jastrow factor, \cite{garnet}  because they potentially provide the input with exact nodes. This improves the single Slater determinant describing a ground-state from the DFT orbitals for heterogeneous systems. DFT nodes may well be poor. The nodes are improved by embedding the MRCI active site.


A simple model of the catalyst surface (planar, no steps, terraces or defects) is retained. This ultimately shows that nano-structuring of the real surface does not prevent an excellent agreement between the apparent activation energy and the QMC calculated value. The energy is referenced to that of adsorbed CO, in its optimised geometry. It is based on a transition state for water addition without complete dissociation. CASINO software \cite{cas2020} is used but cannot be treated as a 'black box'. Many choices need user guidance.  A key step (in fact rate-limiting) of an industrially important reaction, i.e. water-gas shift, focussing on the initial nucleophilic addition of water to pre-adsorbed CO is investigated. This electrophilic carbon is above the center of an equilateral triangle of (111) face Pt-atoms.

These 3 atoms can even be represented as a single unique Pt, the other two being related by translational symmetry. Such a simplification is the key to embedding a high-level wave function for a relatively small active site containing only the two 'distinct' Pt-atoms. One is binding to the H-atom of the stretched water bond and defines our origin and the other defines the triangle within which the hollow site adsorbing CO is situated.

The transition state (TS) geometry obtained (as a saddle point, or energy maximum on this reaction path step) using QMC force constants shows the water oxygen binding to (CO) carbon, whilst the hydrogen from the stretched O-H is bound to a vicinal Pt-atom. The differences in activation energy are significant. For water dissociation on Pt(111), our previous QMC calculation gave 74.15 (with standard error [se] 1.3) kJ/mol compared to a DFT value of 75.2. The present TS (on a thin film) gives a barrier of 64.8 (se 1.5) kJ/mol. There is also evidence that 'free OH' radicals would first react with a surface platinum atom. Hence, diffusion of these 'bound' hydroxides is much more difficult \cite{mukin}.

The standard error gives error-bars of these activation barriers that are much lower than in our QMC benchmarks for Cu(111) \cite{kdd2} . The copper surface has barriers for hydrogen dissociation with standard error estimates slightly above chemical accuracy (1 kcal/mol). This prevents the two mechanistic steps from being distinguished by such barrier differences.  On Pt(111) they are clearly distinguished using QMC calculations. There were some indications on possible improvements to the work on copper \cite{kdd2}. These studies did, nevertheless, provide useful indications on the tendencies. Extrapolation to the basis set limit in plane-waves indicates that the QMC barrier is an upper bound on the actual value and, generally it was rather lower than accurate molecular beam measurements. This tendency explains why the measured value (which may be lower due to defect sites for reaction) agrees with the QMC value well (that has been seen to give values below those measured). In fact, 95\% confidence limits are within 1.6 kJ/mol and this shows a modest effect of random error.

Hence, QMC methods described here represent a breakthrough in reliable information on bond-dissociation limitation of metal catalysed reactions.

QMC was bench-marked for hydrogen dissociation on copper, for which accurate molecular beam measurements are available \cite{kdd2}.


This work is among the largest heterogeneous systems studied by QMC.
The water-gas shift reaction is an equilibrium with the forward direction greatly facilitated on a platinum catalyst producing hydrogen.  We determine the rate-limiting step of the water addition and the associated activation barrier, for CO on Pt(111).

Platinum is a 5d$^9$ 6s$^1$ ground state. All-electron calculations are inaccessible, so each Pt-atom is represented by 10 (or, better, by 18) valence electrons outside a suitable pseudo-potential, noting that little core-valence occupation occurs in any given region of space, because the related densities are well-separated. This confers low variance on the metal wave-function, an order of magnitude lower than that of the (problematic) copper metal wave-functions. The pseudo-potentials discussed below give reliable excitation energies \cite{Raj1}.



Basis and pseudo-potential choice must be carefully made for the present applications. Plane waves with complex exponent allow basis sets to include Slater type orbitals, which facilitates embedding \cite{grun, hog1}.

In early work, a fixed geometry was used, since optimisation was prohibitively long for large systems. No direct experimental geometry for TS species is available and use of the reaction intermediate structure may not be adequate. DFT TS structure was previously input, however, in the present work
the CASINO code is used, its forces algorithm \cite{nemec} gave a molecular active site Slater type orbital TS, expanded in plane-waves like the periodic solid (extending our previous work \cite{hog1}).

Transferrable atomic core electrons must be replaced by effective-core potentials, or pseudo-potentials. Note, however, that these core-electrons account for the majority of the total correlation energy \cite{clem}. The core-size must be chosen small enough to leave all electrons influenced by interactions among the valence electrons (treated explicitly).

The core must be obtained at the Dirac-Fock level (to include scalar relativistic effects, notable for low-energy electrons) and include non-linear core correction (first evidenced by Louie) in order that the pseudo-potential can account for the total core-valence interaction \cite{nlcc}. After these choices, non-local effects may also require special treatment (see below). With these provisos, electron correlation contributions approach the all-electron limit.

This is described in an extensive study which shows that non-locality can cause havoc, if the core is too large, the local channel (l-value) badly defined especially for 3d electrons \cite{kdd1}.

The 4d and 5d are around 10 times lower in variance, ascribed to the lower dispersion of the electron distributions, for increasing effective atomic charge.

\section{The Quantum Monte Carlo Method: specifications.}

\hskip4mm This work evaluates reaction barrier heights. The transition-state geometry is optimised using QMC. The experimental pre-adsorbed CO and water geometry (8 \AA \hskip1mm from the defect-free surface) is used. Subtracting the corresponding energies eliminates most of the non-locality and fixed node non-systematic errors \cite{kdd1}.

The present study simulates heterogeneous catalysis that enhances bond dissociation. This step is frequently the initial (and often limiting) step of an industrial reaction (see, for example \cite{kroes}). Bond dissociation is difficult to describe using most quantum theory approaches, even for isolated diatomic molecules. Taking electron correlation into account does lead to the prediction of the observed products.

Therefore, heterogeneous catalysis requires an approach correctly giving almost all electron correlation (the related energy varies as bonds dissociate), in addition to the interactions involved.
Hartree-Fock methods incorrectly describe bond dissociation. This is so, even for homo-nuclear diatomics. Extensive CI caters for electron correlation adequately. The DFT methods are a rapid computational alternative, with some correlation which performs better in the dissociation limit for certain functionals.

The Perdew, Burke, Ernzerhof (PBE) functional is used since it previously gave accurate lattice parameter values. It gives a reasonable gas phase barrier for ammonia synthesis, however, on Pt(111) limiting the relaxed compact cubic structure, the barrier increased significantly, as opposed to observed catalytic effects \cite{honk}.

QMC which includes correlation explicitly is required. The trial wave-function must behave as correctly as possible, in particular close to dissociation. A high-level wave-function embedded in periodic  Kohn-Sham PBE plane-waves is a suitable starting point for trial wave-functions in Quantum Monte Carlo work on transition metal systems (see below).

 The correlation energy contribution varies dramatically during adsorption and reaction, therefore it must be determined exactly.  The accuracy of this term in QMC is uniquely high. A set of DFT benchmarks by Thakkar \cite{thakk} for evaluating electron correlation energy highlights the poor performance of 11 much used DFT functionals. A comparative study of Diffusion Monte Carlo (DMC) and DFT-MRCI (DFT-Multi-reference configuration interaction) methods for excitation energies tests similar cases to the present rate-limiting activation barriers (in system size and electron re-arrangement): percentage errors in the total excitation energies are 3 \% for DFT-MRCI and only 0.4 \% for DMC. (See Lester \cite{les}).

After equilibration, for many data-points N, the error decreasing as 1/$\sqrt{N}$.

Trial wave-function quality is carefully optimised and finite size effects catered for.  Solid-state QMC can be made to scale slowly with system size (n electrons scale as n$^3$), expanding the plane-wave basis in cubic splines (blips) \cite{alfe}.
The CASINO code is used, which is well-suited to periodic solids.

The purpose of the work reported here is to study the stabler of two possible catalytic reaction paths for water addition to pre-adsorbed CO, distinguished via their specific TS.

The statistical error (0.7 kJ/mol) within which the calculated reaction barrier heights are located, must be taken as including specific non-systematic contributions (due to non-locality of the pseudo-potential and poor nodes of the trial wave-function).

We found water addition to CO pre-adsorbed on Pt (111) can occur in two ways: \cite{absi}

I-a step by step process, with rate limiting water dissociation on Pt \cite{phat}.

II-a concerted step, with CO and water co-ordinated to the metal.

This work studies the concerted addition of water (II): QMC energies of asymptotic physisorbed geometries for CO and H$_2$O are subtracted from that of the stringently optimised QMC geometry for the adsorbed Transition State (TS).

The TS geometry uses QMC force constants (the Hessian is defined with QMC force constants as matrix elements, then updated using Pulay's Direct Inversion in the Iterative Subspace (DIIS) algorithm adapted by Farkas \cite{fark}). The TS is shown in Figure 1. It has a slightly short CO bond, because some electron density involved in beginning the bonding with the water oxygen comes from a CO anti-bonding orbital. The water molecule has a very long OH-bond (1.2 \AA), its hydrogen binding to Pt at (0,0,0). The other OH is almost unchanged.

In this study, we have established its geometry using optimisation guided by QMC force constants and precisely estimated the activation barrier height.

\subsection{Setting up the model system.}

\hskip4mm These systems involve stretched bonds requiring almost all the electronic correlation. For such geometries, our previous work has shown \cite{boufh} that QMC captures much more correlation energy than the majority PBE-DFT value. QMC is therefore reliable for near-exact correlation in bond-dissociation, even more so than correlated Hartree-Fock based methods.
\vskip4mm
For heterogeneous systems involving solids, the Generic Jastrow factor (see below) should be used, accounting for translational symmetry.
For this application to catalysis, a slab of the metal is constructed. This is repeated periodically in both directions (x and y) within the solid and defining a planar (111) surface at z=0. Periodicity in the z-direction is introduced by repeating the slab after a large vacuum spacing (20.8 \AA) and is limited to the half-axis 'outside' the catalyst that includes the adsorbed molecules, with the bulk z-axis retaining platinum periodicity perpendicular to the Pt(111) face. The x and y dimensions should be much longer than the maximum bondlength and sufficiently thick to absorb the surface layer perturbation.
The 5-layer Pt-ABABA slab top 3 layers are spaced according to experimental measurement. The last two use exact bulk a = 3.912 \AA. In plane Pt(111) atom-spacing is $ a \over \sqrt{2}$ and the last layers are $ a \over \sqrt{3}$ apart to ensure continuity with the bulk.

The surfaces in this model are planar, whereas it is known that metal close-packed faces re-arrange forming high Miller index ridges or meanders. \cite{pimp} Our planar-surface model uses the following states:
The reference state with the same atoms as the TS (i.e. just before the reaction): (1) The physisorbed asymptote for distant H$_2$O. The water molecule is placed 8 \AA \hskip1mm from the slab with pre-adsorbed CO. A double cell run per molecule is needed.

(2) Co-adsorbed molecules are 2 \AA \hskip1mm above the Pt(111) plane but 8 \AA \hskip1mm apart.

(3) The Transition-state (TS) geometry (see Figure 1).

This geometry 3 serves to initialize the QMC transition-state.

To limit finite-size effects, a 3 3 1 grid defines the largest real-space super-cell, much smaller than the converged DFT/plane wave grid (limited by computer memory).

The DFT calculations converge only at 16 16 1 or higher. They are useful as control variate in twist-averaging. This corrects for finite size error which cancels between asymptotic and Transition State geometries leaving below 1/10 of the error. Runs on a 2 2 1 grid reduced finite size-error by a factor 25 (127 for 3 3 1) compared to a gamma-point calculation.

The pseudo-potential must allow inclusion of semi-core electrons in the valence. This is certainly necessary when dissociating molecules involving the transition metal atoms. It also improves test results for platinum. The copper 3d$^{10}$ shell is dense in the core-region but platinum has its 5d$^9$ 6s$^1$ ground-state shell on average further from the core. Pt is thus less prone to difficulties defining pseudo-potentials. Semi-core electrons also appear to have much less influence on metallic systems in which the number of metal atoms is conserved and geometry very similar (almost bystanders). Such is the case when comparing asymptote and QMC optimised TS geometries for reaction barriers. Co-ordination of the TS to the metal surface occurs but the distance between atoms is about 40 \% more than the equilibrium bond--length and so the role of semi-core electrons is minor. A Z=60 core for Pt is validated (see below).

A summary of the QMC strategy used is given in our previous work on Pt(111) as a catalyst for this reaction.

The same TS geometry was adopted. A QMC saddle point or energy maximum on this reaction path step gives us TS geometry.

\begin{figure}
\includegraphics[scale=0.07]{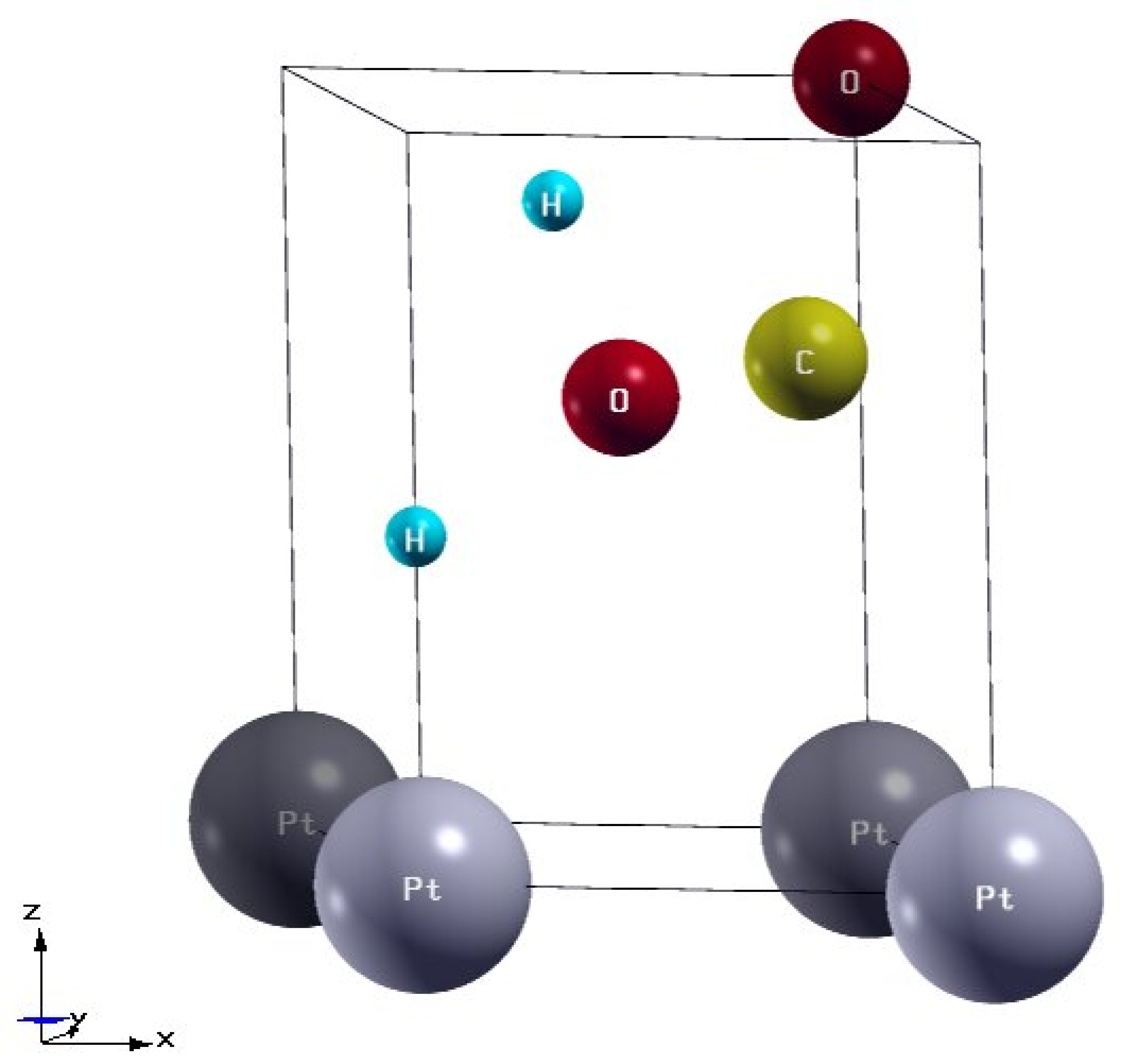}
\includegraphics[scale=0.07]{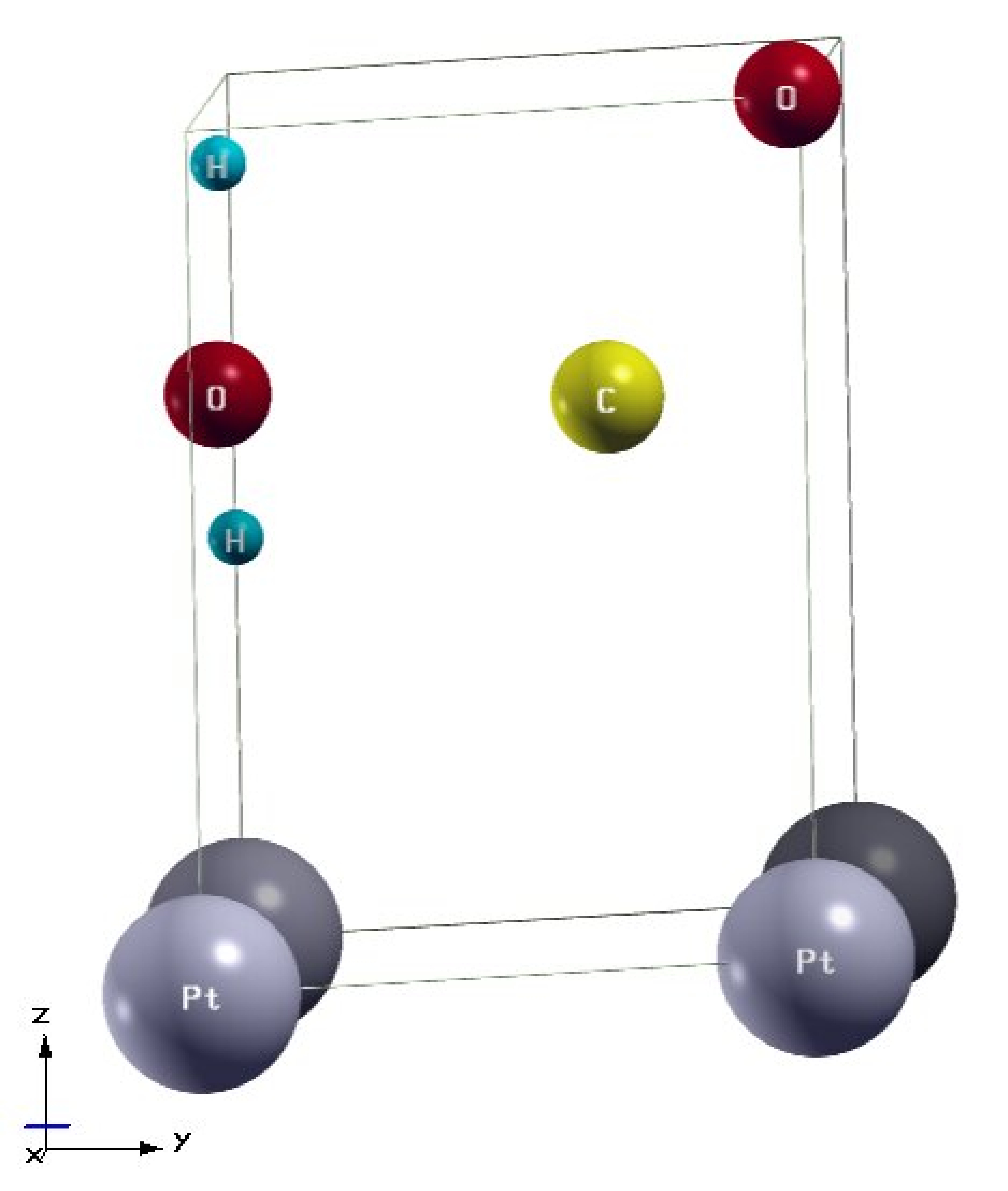}
\caption{Molecular active site (TS) TS1: the z=0 plane is the Pt(111) surface (bottom face). Pt at the origin accepts one water H-atom.  CO carbon in a hollow site (Pt-C bond-length 1.43 \AA), The oxygen is above a vicinal Pt. Water (oxygen) O--C distance is 1.63 \AA. \\ Top pane: from negative y.  \\ Bottom pane: from positive x}
\end{figure}

\section{Results.}
\subsection{Advantages of thin-films.}
The catalyst studied was represented as a platinum monolayer on Al(111). This interface has a lattice miss-match, which results in stretching the Pt--Pt linkages in the monolayer by 3.5 \%. This, in turn, activates them, both towards CO adsorption and also results in a more reactive water attack overall, with the activation barrier lowered significantly by 6 kJ/mol.
\subsection{Validating a pseudo-potential for simple reactions of platinum:}
\hskip4mm Previous work used Troullier Martins norm conserving pseudo-potentials (TMPP), which were shown to be inadequate in terms of accuracy for the test reaction below, giving PtH.

The present work therefore converts the Effective Core Potential (ECP) used in Molpro.

It is limited to the active site molecular wave-function and in CASINO we use throughout a norm conserving form which has the same local channel (l=1) as ECP60MDF.

The reaction Pt$_2$ + H$_2$ $\rightarrow$  2PtH also tests the performance of the options at DMC (and, most importantly the PP).

The integrals over PP (including the Casula T-move correction for non-locality) are numerical with truncated series (some loss of accuracy).
The TMPP is relatively hard, conferring low variance on the wave-function.


As previously discussed, pseudo-potentials for Pt are constructed so that less QMC variance is obtained than 3d metals, like copper \cite{kdd2}.

The 18-electron core is required for chemical accuracy in this test reaction over the 10 electron TMPP which is not sufficiently accurate.

\vskip1mm
Pt + H$_2$ (giving PtH$_2$ (a) and Pt$_2$ + H$_2$ giving 2 Pt-H (b).
\vskip1mm
With the TMPP, QMC reaction energy for (b) that was shown to proceed via a squarish/rectangular TS is around 9 kJ/mol above reference values (MRSDCI calculations \cite{balb}).

For Pt$_2$ + H$_2$   binding energy of 137.1 kJ/mol is predicted by QMC (c.f. 128.3 kJ/mol obtained from MRSDCI benchmarks. This is ascribed to local approximation errors.

We designed a hard pseudo-potential (with 18 electrons treated explicitly) which is based on the MDF60 PP of the Stuttgart group \cite{fpseu} (which treats as valence 9*5d 1*6s and 8*4f, most notably retaining 6 of the 4f electrons in core) and found it allowed us to accurately simulate the test reaction molecules. The QMC binding energy was 130 kJ/mol (c.f. 128.3) \cite{balb} with a standard error of 2 kJ/mol.

These results show our approach is valid, using input from Molpro as trial wave-function for molecular systems. The final step is embedding these reacting systems in a periodic cell.

\hskip4mm  The water-gas shift reaction on Pt(111) is a source of  sustainable H$_2$ production. This reaction of carbon monoxide and water is practically unknown in the gas phase. We use QMC to describe CO+H$_2$O adsorbed on a Pt(111) surface. The surface has 2-D periodicity. Molecules interact with both the surface and each other. A slab construct, limited by relaxed Pt(111) (reaching the experimental bulk geometry in 4 layers) is used. Carbon monoxide (CO) adsorption is considered. Carbon monoxide gas is rather inert and the carbon atom even carries a small partial negative charge.

Upon adsorption (after physisorption that little alters CO), charge transfer to a Pt-surface restores the typical reactive carbonyl species, with a positively charged carbon that is the site of nucleophilic attack  (with a partial charge 0.098 e). The strong bonding within this molecule when in the gas phase is weakened by interactions with the surface, making the molecule easier to attack by water etc.

The adsorbed carbon monoxide molecule is polarized like a carbonyl moiety, after charge transfer to the surface. It is then reactive towards nucleophiles.

Investigation of this phenomenon has been carried out using infrared reactor cells (in situ FTIR, \cite{baz}) at a number of solid surfaces. Here, for Pt(111), it is complemented by a QMC study of the reaction with water. Removal of toxic carbon monoxide molecules is a typical de-pollution reaction, of interest in catalytic exhausts. Furthermore, the reaction with water is of industrial importance in producing clean fuels (hydrogen gas) in a sustainable process.
In earlier work \cite{hog2,baz},it was shown that the stretch frequency of carbon monoxide makes it a highly sensitive probe to surface interactions. This frequency is a measure of adsorption. The carbon monoxide group also acquires the {\bf carbonyl} polarization, i.e. is reversed to become the site of nucleophilic attack.

{\bf CO and water:}

Water reacts with pre-adsorbed CO, as shown by preliminary investigations using {\it ab initio} DFT (PBE functional) studies, with a plane-wave basis. The reaction products are CO$_2$ and gaseous hydrogen (fuel restoring water as combustion product). CO adsorbed on Pt (111) was optimized using QMC force constants and found to be 118 pm, only slightly longer than the gas-phase value of 113 pm which is partly due to the interaction with water depleting electron density of a CO anti-bonding orbital). This CO can be studied by FTIR and shown to possess a partial positive charge on carbon, suitable for nucleophilic attack. Density analysis of QMC results shows the partial charge is +0.098 e at a Pt-C distance of 1.82 \AA.
Water dissociation concerted with oxygen beginning to link with the CO carbon is rate limiting. Dissociation preceding reaction is less favorable but was the reaction channel initially studied.
In Figure 1, (see the above {\bf model} section) we show the present QMC geometry of the TS. The water molecule with the leaving H$^{\delta+}$ binds with a surface Pt-atom, whilst the remaining O-H is forms a V-shaped H--O$^{\delta-}$--C$^{\delta+}$=O intermediate. The carbon-monoxide carbon atom and water oxygen are 2.27 \AA \hskip1mm above the Pt(111) face. The resulting QMC optimized geometry determines three Pt-O tripod distances as 2.04 \AA.

\vskip4mm
The O-C linkage forming is still 1.63 \AA \hskip1mm long (in this TS).
The four-Pt atom {\bf skin} lozenge are at the apices of a trigonal (i.e. hexagonal) close-packed cell with Pt-Pt distance 2.783 \AA.


Water dissociation concerted with the start of nucleophilic attack of CO at Pt(111): QMC Study based on the TS of Figure 1 above, in the Model section (with the carbonyl polarization of adsorbed CO, i.e. partial positive charge on carbon unlike CO (gas)).

textcolor{blue} {Tabulated values are for the DMC population target weight.
\vskip1mm
 \begin{tabular}{|c|c|c|c|}
  \hline
 Structure & Transition-state  & Asymptote &  Water\\
 \hline
 E$_{tot}$ (Ha) & -182.232285 & -165.07455 & -17.1824 \\
\hline
Variance au/(kJ/mol)$^2$ : se kJ/mol& 0.507/3.5: 1.87* & 0.528/3.64: 1.91* & 0.3 \\
\hline
\end{tabular}}
\vskip2mm
{\bf Table I:Activation energy evaluation for online data-sets from this work \cite{bcs}}
\vskip1mm
Activation barrier: 0.02466 au or 64.77 (5) $\pm$ 1.5 kJ/mol, (using *) showing comparable standard error value in kJ/mol for TS and asymptotic gamma point distributions, which validates our MRCI input.

The final QMC barrier of 64.8 with a standard error of 1.5 kcal/mol. This QMC value could be taken as a lower bound for the activation barrier, limited by the activated complex. The actual system will depend on temperature, surface re-arrangement and possible defects. For the pure Pt(111) face: apparent activation energy measured is 71.05 kcal/mol.

\vskip4mm
\section{Perspectives and conclusions.}

\hskip4mm This Quantum Monte Carlo determination of the Transition state structure and reaction barrier height for water CO addition provides new insight into producing hydrogen as a sustainable energy source, with an industrial catalyst. The concerted mechanism is favored.

This work shows that water dissociation followed by addition of hydroxide to pre-adsorbed CO has an activation barrier 1 kcal/mol higher as well as OH adsorption that impairs it.

The catalyst studied is platinum, acting via its (111) close-packed surface. A small molecular structure, described by Slater type atomic orbitals is optimized using QMC forces and the CASINO software. This geometry is used to initialize plane-wave DFT wave-function evaluation providing Kohn-Sham PBE orbitals to define the trial QMC wave-function completed by a complex Jastrow factor. This embedding procedure is validated for some test reactions of hydrogen on platinum clusters. The Pt(111) slab used here was also tested.
The barrier height is therefore not expected to suffer from systematic errors and the statistical (standard) error of 0.2 kcal/mol is realistic, and comparable to that of measured apparent activation energy.

Perspectives for extending this strategy come from full-CI input for molecules available in the Alavi group. At the time of writing, this code provides QMC force constants for all-electron calculations on molecules. The method needs to be tested for effective core potentials (ECP) and the geometry embedding for plane-wave input needs to be bench-marked. This is the aim of one of our current research projects.



We refer to CO adsorbed above the centroid of an equilateral triangle with Pt-atoms as apices (defined using translational symmetry from a single unique Pt-atom) in the Pt(111) plane. This gives three C-Pt linkages.
The data giving the TS structure uses CASINO QMC forces code and a contact molecule, comprising CO+H$_2$O and a monolayer of Pt(111) optimized using QMC and placed on the slab at the experimental spacing. The size and complexity of this system require considerable supercomputer resources.

{\bf Acknowledgements.}


The QMC calculations were made possible by allocation of supercomputer resources to DARI project spe00018: 22.6 Million core-hours on the Irene supercomputer (CEA, Bruy\`eres-le-Ch\^atel), near Paris, France.



\providecommand{\latin}[1]{#1}
\makeatletter
\providecommand{\doi}
  {\begingroup\let\do\@makeother\dospecials
  \catcode`\{=1 \catcode`\}=2 \doi@aux}
\providecommand{\doi@aux}[1]{\endgroup\texttt{#1}}
\makeatother
\providecommand*\mcitethebibliography{\thebibliography}
\csname @ifundefined\endcsname{endmcitethebibliography}
  {\let\endmcitethebibliography\endthebibliography}{}
\begin{mcitethebibliography}{0}
\providecommand*\natexlab[1]{#1}
\providecommand*\mciteSetBstSublistMode[1]{}
\providecommand*\mciteSetBstMaxWidthForm[2]{}
\providecommand*\mciteBstWouldAddEndPuncttrue
  {\def\EndOfBibitem{\unskip.}}
\providecommand*\mciteBstWouldAddEndPunctfalse
  {\let\EndOfBibitem\relax}
\providecommand*\mciteSetBstMidEndSepPunct[3]{}
\providecommand*\mciteSetBstSublistLabelBeginEnd[3]{}
\providecommand*\EndOfBibitem{}
\mciteSetBstSublistMode{f}
\mciteSetBstMaxWidthForm{subitem}{(\alph{mcitesubitemcount})}
\mciteSetBstSublistLabelBeginEnd
  {\mcitemaxwidthsubitemform\space}
  {\relax}
  {\relax}

\end{mcitethebibliography}


\begin{thebibliography}{90}

\bibitem{shar} Selective hydrogen production at Pt(111) investigated by Quantum Monte Carlo methods for metal catalysis. Sharma, R.O., Rantala, T.T. and Hoggan, P.E. Int. J. Quantum Chem. (2020) https://onlinelibrary.wiley.com/doi/abs/10.1002/qua.26198
\bibitem{hjcp} Sharma, R.O.; Rantala, T.T.; Hoggan, P.E. Quantum Monte Carlo approach for determining the activation barrier of water addition to carbon-monoxide adsorbed on Pt(111) within 1 kJ/mol. {\it J. Phys. Chem C.} {\bf 2020 } 124, 26232-26240.
\bibitem{kdd3}     Powell, A.D.; Kroes, G.J.; Doblhoff-Dier, K. Quantum Monte Carlo calculations on dissociative chemisorption of H2 + Al(110): minimum barrier heights and their comparison to DFT values. {\it J. Chem. Phys.} {\bf 2020} 153, 224701.
\bibitem{oakr} Krogel, J. T.; Yuk, S. F.; Kent, P. R.C.; Cooper, V. R. Perspectives on van der Waals density functionals: the case of TiS2. {\it J. Phys. Chem A} {\bf 2020} 124, 47, 9867--9876.
\bibitem{hoggadv} Hoggan, P. E. Quantum Monte Carlo with ground-state input to investigate platinum doped aluminium catalyst: H$_2$ production from adsorbed CO.  {\it Adv. Quantum Chem.} {\bf 2021} 83, 155-170.
\bibitem{geo} Rinaldi, D: Cartier, A; Hoggan, P. E. GEOMOS: semi-empirical SCF system for dealing with solvent effects and solide surface adsorption. QCPE Bull, {\bf 1989} 9, 128.
\bibitem{mavr} On the Mechanism of Low-Temperature Water Gas Shift, Reaction on Copper, A.  A. Gokhale, J. A. Dumesic, and M. Mavrikakis, J. Am. Chem. Soc. 2008,130, 1402-1414
\bibitem{mukin} Grabow, L. C.; Gokhale, A. A.; Evans, S. T.; Dumesic, J. A.;  Mavrikakis, M., Mechanism of the Water Gas Shift Reaction on Pt: First Principles, Experiments, and Microkinetic Modeling. J. Phys. Chem. C. 2008,112 4608-4617;
\bibitem{Camp} Campbell, C. T., Energies of Adsorbed Catalytic Intermediates on Transition Metal
Surfaces: Calorimetric Measurements and Benchmarks for Theory. Acc. Chem. Res.2019, 52, 984--993.
\bibitem{Clay} Clay, J. P. ; Greeley, J. P.; Ribeiro, F. H. ; Delgass,  W. N.; Schneider,  W. F.; DFT comparison of intrinsic WGS kinetics over Pd and Pt.  J. of Catal. 320 (2014) 106--117.
\bibitem{Faj2020} Fajin, J. L.C. ;  Cordeiro, M. N. D.S., Probing the efficiency of platinum nanotubes for the H2production by water
gas shift reaction: A DFT study. Applied Catalysis B: Environmental 263 (2020) 118301.
\bibitem{cas2020} R. J. Needs, M. D. Towler, N. D. Drummond, P. Lopez Rios, J. R. Trail. Variational and Diffusion Quantum Monte Carlo Calculations with the CASINO Code. J. Chem. Phys. 152, 154106 (2020). URL: https://doi.org/10.1063/1.5144288
\bibitem{kdd2} "Quantum Monte Carlo calculations on a benchmark molecule - metal surface reaction: H2 + Cu(111)", K. Doblhoff-Dier, J. Meyer, P. E. Hoggan and G-J Kroes. J. Chem. Theory  Comput.  2017, 13, 7, 3208-3219  DOI: 10.1021/acs.jctc.7b00344 (Published online 17 may 2017.
\bibitem{bcs} Figshare DOI for Raw QMC data corresponding to Table I here: 10.6084/m9.figshare.10293194
\bibitem{Raj1} Rajesh O. Sharma and Philip E. Hoggan, Physisorption energy of H and H2 on clean Pt(111) as a useful surface energy reference in Quantum Monte Carlo calculation, Advances in quantum chemistry 79 (2019) p311-322
\bibitem{grun} Grueneis, A.; Shepherd, J. J.;  Alavi, A.; Tew, D. P.; Booth,  G. H.  Explicitly correlated plane waves: Accelerating convergence in periodic wave-function expansions. J. Chem. Phys. 2013, 139, 084112.
\bibitem{hog1} Hoggan, P. E.; Quantum Monte Carlo simulation of carbon monoxide reactivity when adsorbed at metal and oxide catalyst surfaces: Trial wave-functions with exponential type basis and quasi-exact three-body correlation. Int. J. Quantum Chem. 2013, 113,  277-285.
\bibitem{hog2} Hoggan, P. E.; Bouferguene, A. Relative Advantages of Quantum Monte Carlo Simulation for Changing Electron Correlation: CO Reactions on Copper and Platinum Catalysts. Adv. Quantum Chem. 2014, 68, 89-104.
\bibitem{balb} Balasubramanian, K.; Potential energy surfaces for the Pt2+H2 reaction. J. Chem. Phys.  1991, 94, 1253-1263.
\bibitem{kroes} Kroes, G. J. Towards chemically accurate simulation of molecule–surface reactions. Phys. Chem. Chem. Phys. 2012, 14, 14966-14981.
\bibitem{diaz} Diaz, C.; Pijper,  E.;  Olsen, R. A.; Busnengo,  H. F.; Auerbach,  D. J.;  Kroes, G. J. Chemically Accurate Simulation of a Prototypical Surface Reaction: H2 Dissociation on Cu(111). Science. 2009, 326, 832-834.
\bibitem{nemec} N. Nemec, M. D. Towler and R. J. Needs. J. Chem. Phys. 132, 034111 (2010); https://doi.org/10.1063/1.3288054
\bibitem{clem} Clementi E and Corongiu G., Note on the atomic correlation energy.  Int J Quant Chem 62: 571-591, 1997
\bibitem{nlcc} Louie, S. G.; Froyen, S.; Cohen, M. L. "Nonlinear ionic pseudopotentials in spin-density-functional calculations", Phys. Rev. B, 26, 1738-1742 (1982).
\bibitem{kdd1} Diffusion Monte Carlo for accurate dissociation energies of 3d transition metal containing molecules
K. Doblhoff-Dier, J. Meyer, P. E. Hoggan, G-J Kroes, and L. K Wagner. J. Chem. Theory  Comput. 2016, 12, 2583-2597.
Publication Date (Web): May 13, 2016 (Article) DOI: 10.1021/acs.jctc.6b00160
\bibitem{garnet}  Changlani, H. J.; Kinder, J. M.; Umrigar, C. J.; Chan, G. K-L. Approximating strongly correlated wave functions with correlator product states. Phys. Rev. B. 2009, 80, 245116
\bibitem{asy} Drummond, N. D.; Needs, R. J.; Sorour, A.; Foulkes, W. M. C. Finite-size errors in continuum quantum Monte Carlo calculations.
Phys. Rev. B. 2008, 78, 125106.
\bibitem{con} Needs, R. J.; Towler, M. D.; Drummond, N. D.; Lopez Rios, P, J. Continuum variational and diffusion quantum Monte Carlo calculations. Phys. Condens. Matter. 2010, 22, 023201.
\bibitem{wfq} Reinhardt, P.;  Hoggan, P. E. Cusps and derivatives for H2O wave-functions using Hartree-Fock Slater code: a density study.  Int. J. Quantum Chem. 2009, 109, 3191-3198.
\bibitem{tlse} Toulouse, J.; Hoggan, P. E.; Reinhardt, P.; Caffarel, M.; Umrigar, C. J.  Quantum Monte Carlo Calculations of Electronic Excitation Energies: The Case of the Singlet n-$\pi$(CO) Transition in Acrolein. Prog. Theo. Chem. Phys. B. 2012,  22,  343-351.
\bibitem{jas} Drummond, N. D.; Towler, M. D.; Needs, R. J. Jastrow correlation factor for atoms, molecules, and solids. Phys. Rev. B. 2004, 70, 235119
\bibitem{honk} Honkala, K.; Hellman, A.; Remediakis, I. N.; Logadottir, A.; Carlsson, A.; Dahl, S.; Cristensen, C. H.; Norskov, J. K. Ammonia Synthesis from First-Principles Calculations. Science. 2005, 307, 555-558.
\bibitem{thakk} Thakkar, A. J.; McCarthy, S. P. Towards improved density functionals for correlationn energy. J. Chem. Phys. 2009, 131, 134109.
\bibitem{les} Lester, Jr, W. A.; Mitas, L.; Hammond, B. Quantum Monte Carlo for atoms, molecules and solids. Chem. Phys. Lett.  2009, 478, 1-10.
\bibitem{alfe} Alf\`e, D.; Gillan, M. J. Efficient localized basis set for quantum Monte Carlo calculations on condensed matter.  Phys. Rev. B. 2004, 70, 161101.
\bibitem{absi} N. Absi and P. E. Hoggan, Quantum Monte Carlo investigation of two catalytic reaction paths for hydrogen synthesis on Pt(111). In Recent progress in Quantum Monte Carlo, ed. S.Tanaka, L. Mitas, P-O Roy, ASC books, Symposium Series 2016, 1234 Chapter 5, 77-88.
\bibitem{phat} A. A. Phatak, A. A.; Delgass, W. N.; Ribeiro, F. H.; Schneider, W. F., Density Functional Theory comparison of water dissociation steps on Cu, Au, Ni, Pd and Pt. J. Phys Chem C. 2009, 113 7269-7276.
\bibitem{fark} O. Farkas and H. B. Schlegel, Phys. Chem. Chem. Phys., 2002, 4, 11-15.
\bibitem{boufh} P. E. Hoggan and A. Bouferguene, Quantum Monte Carlo for Activated Reactions at Solid Surfaces: Time Well Spent on Stretched Bonds, International Journal of Quantum Chemistry, 2014,114 p1150-1156.
\bibitem{pimp} Ben Hadj Hamouda,  A.;   Absi, N.;  Hoggan, P.E.; Pimpinelli,  A.
Growth instabilities and adsorbed impurities: a case study. {\it Phys Rev B} {\bf 2008} 77 245430.
\bibitem{molpro} MOLPRO package of ab initio programs. H.-J. Werner, P. J. Knowles, G. Knizia, F. R. Manby,
\bibitem{fpseu} D. Figgen, K. A. Peterson, M. Dolg and H. Stoll,  Energy-consistent pseudopotentials and correlation consistent basis sets for the 5d elements Hf-Pt. J. Chem. Phys. 2009 130 164108.
\bibitem{neci} NECI. FCIQMC code developed by George Booth and Ali Alavi, 2013. (GNU licence).
\bibitem{bens}P.E. Hoggan, M. Bensitel and J.C.  Lavalley, A new method of calculating interactions between adsorbates and metal oxide surfaces: application to the study of CO2 insertion in hydroxyl or methoxy groups on Al2O3 and TiO2. J. Mol. Struct., 1994, 320, 49-56.
\bibitem{splin} D.  Alf\`e and M. Gillian, Efficient localized basis sets for quantum Monte Carlo calculations on condensed matter. Phys. Rev. B. 2004, 70, 161101.
\bibitem{abin} www.abinit.org
\bibitem{casino}  Needs, R. J.; Towler, M. D.;  Drummond, N. D.;  Lopez Rios,  P. Continuum variational and diffusion quantum Monte Carlo calculations.
 J. Phys.: Condens. Matter 22, 023201 (2010);
casino website: www.tcm.phy.cam.ac.uk/mdt26/casino2.html.
\bibitem{baz} Bazin, P.; Saur, O.; Lavalley, J. C.; Daturi, M.; Blanchard, G. FT-IR study of CO adsorption on Pt/CeO2: characterisation and structural rearrangement of small Pt particles. Phys . Chem. Chem. Phys. 2005 , 7, 181-194
\bibitem{faj} Fajin, J. L. C.; Cordeiro, M. N. D. S.; Gomes, J. R. B. Density Functional Theory Study of the Water Dissociation on Platinum Surfaces: General Trends, J. Phys. Chem. A. 2014, 118, 5832-5840.
\end{thebibliography}
\end{document}
\